\documentclass[prc,preprint,showpacs,superscriptaddress,floatfix,amsmath,amssymb,nofootinbib]{revtex4}
\usepackage{graphicx}
\usepackage{color}
\usepackage{ulem}
\usepackage{multirow}
\usepackage{slashbox}
\usepackage{booktabs}

\newcommand{\beq}{\begin{equation}}
\newcommand{\eeq}{\end{equation}}
\newcommand{\beqn}{\begin{eqnarray}}
\newcommand{\eeqn}{\end{eqnarray}}

\begin{document}
\title{Electromagnetic transitions in multiple chiral doublet bands}
\author{Hui Jia } \author{Bin Qi}\thanks{e-mail: bqi@sdu.edu.cn} \author{Shou-Yu Wang}\author{Shuo Wang} \author{Chen Liu}
\address{Shandong Provincial Key Laboratory of Optical Astronomy and Solar-Terrestrial Environment,
Institute of Space Sciences, Shandong University, Weihai 264209, China}

\date{\today}

\begin{abstract}
Multiple chiral doublet bands (M$\chi$D) in the $80$, 130 and $190$ mass regions are studied by the model of  $\gamma$=90$^{\circ}$ triaxial rotor coupled with identical symmetric proton-neutron configurations. By selecting the suitable basis, the calculated wave functions are explicitly exhibited to be symmetric  under the operator $\hat{A}$, which is defined as rotation by $90^{\circ}$ about 3-axis with the exchange of valance proton and neutron. We found that both $M1$ and $E2$ transitions are allowed between the levels with different values of $A$, while are forbidden between the levels with same values of $A$.
Such a selection rule holds true for M$\chi$D in different mass regions.
\end{abstract}


\maketitle

\section{Introduction}
Chirality in nuclei was originally predicted  by Frauendorf and Meng~\cite{FM97} in 1997 for triaxially
deformed nuclei, which exhibited as a pair of nearly  degenerate $\Delta I = 1$ bands with the same parity, namely chiral
doublet bands. Chiral doublet bands were first observed in 2001 in the $N$$=$75 isotones~\cite{Starosta01}. Later, both theoretical and experimental
effort has been devoted to search for more chiral nuclei. So far, more than 30 candidate chiral nuclei have been reported experimentally in the 80, 100, 130 and 190 mass regions \emph{}\cite{Koike01,Bark01,Hecht01,Hartley01,Mergel02,Koike03,Zhu03,Vaman04,Joshi04,Timar04,Alcantara04,WangSY06a,Grodner06,Tonev06,Timar06,Joshi07,Lawrie08,Suzuki08,Wang11,Koike05,Masiteng13,Liu16}.
Theoretically, nuclear chirality has been investigated in
the frameworks of the titled axis cranking approch~\cite{FM97,Dimitrov00PRL,Olbratowski04,Mukhopadhyay07}, particle rotor model (PRM)~\cite{FM97,PengJ03,Zhang07,WangSY08,Qi0901,Qi12,Wang08,Wang09,Liul13,Wangsy07,Wangsy10,Meng08}, interacting boson approximation (IBA)~\cite{Tonev06,Tonev07} and project shell model~\cite{Bhat14}.

In 2006, based on adiabatic and configuration fixed constrained triaxial covariant density functional theory, Meng $et~al.$~\cite{MengJ06} predicted that multiple chiral doublet bands (M$\chi$D) with different deformations and different intrinsic configurations could exist in one single nucleus. The first experimental evidence for M$\chi$D was obtained in $^{133}$Ce~\cite{Ayangeakaa13} with the configurations $\pi g_{7/2}^{-1}h_{11/2}^{1} \otimes \nu h_{11/2}^{-1} $ and $\pi h_{11/2}^{2} \otimes \nu h_{11/2}^{-1} $, then M$\chi$D with different configurations were suggested in $^{107}$Ag~\cite{Qi13} and  $^{78}$Br~\cite{Liu16}.
In addition, a novel type of M$\chi$D with identical intrinsic configurations was also theoretically discussed in the $130$ mass region with the configuration $\pi h_{11/2} \otimes \nu h^{-1}_{11/2}$~\cite{Droste09,Qb10,Ikuko13,Chen16} and in the $100$ mass region with the configuration $\pi g_{9/2}^{-1} \otimes \nu h_{11/2}$~\cite{Ikuko13}. Recently, this new type of M$\chi$D was experimentally reported in $^{103}$Rh~\cite{Kuti14}, which was analyzed by using the tilted axis cranking covariant density functional theory~\cite{Peng08,Peng15,Zhao11,Zhao111,Zhao12,Meng13} along with PRM. Such new type of M$\chi$D have not been discussed in the 80 and 190 mass regions so far.

Besides the degeneracy of excitation energies, the properties of the electromagnetic transitions are considered as another criteria for confirming chiral doublet bands. In 2004, Koike $et~al.$~\cite{Koike04} introduced the selection rule  for electromagnetic transition probabilities.
In an ideal case of a $\gamma=90^{\circ}$ rotor coupled to a symmetric particle-hole configuration, a new operator $\hat{A}$, which is defined as rotation by $90^{\circ}$ about the 3-axis with the exchange of valance proton and neutron, was used to represent the chiral operator. The selection rule in terms of the quantum number $A$ was examined by numerical calculations for the lowest chiral doublet bands with the configuration $\pi h_{11/2} \otimes \nu h^{-1}_{11/2}$ in the 130 mass region~\cite{Koike04}.
Then the selection rule was used to analyse the excited chiral doublet bands in the 130 mass region~\cite{Ikuko13}.  It is easy to imagine that the selection rule for electromagnetic transitions should be directly connected to the symmetry of the wave functions. However, explicit expressions for the wave function with symmetry under $\hat{A}$ have not yet been shown for the chiral doublet bands.

These facts motivate us to make more detailed theoretical studies for the electromagnetic transitions of M$\chi$D  in the different mass regions. In this paper, by adopting the particle rotor model, we systematically study the symmetry of wave functions and the selection rule for M$\chi$D with $\gamma$=90$^{\circ}$ rotor coupled with the identical intrinsic configuration, i.e., $\pi g_{9/2} \otimes \nu g^{-1}_{9/2}$,  $\pi h_{11/2} \otimes \nu h^{-1}_{11/2}$ and $\pi i_{13/2} \otimes \nu i^{-1}_{13/2}$ for the 80, 130 and 190 mass regions, respectively.

\section{Formalism}

The total Hamiltonian of particle rotor model of odd-odd nuclei can be written as~\cite{FM97,Koike04,Qi0902}
\begin{equation}\label{H1}
    \hat{H} = \hat{H}_{\rm core}+\hat{H}_{\rm p}+\hat{H}_{\rm n}.
\end{equation}
The Hamiltonian of the core is
\begin{equation}\label{}
   \hat{H}_{{\rm core}} = \sum^{3}_{k=1}\dfrac{\hat{R}^{2}_{k}}{2{\cal J}_{k}},
\end{equation}
where the indices $k$ = 1, 2, 3 represent three principal axes of the body-fixed frame, and $\hat{R}_{k}$ represents the angular momentum operators for core, and ${\cal J}_{k}$ represents moment of inertia for irrotational flow, i.e.,
\begin{equation}\label{}
  {\cal J}_{k} = {\cal J}_{0}\sin^{2}(\gamma-2\pi k/3)~~~~~~~k=1, 2, 3.
\end{equation}
For $\gamma$=90$^{\circ}$, ${\cal J}_{1}=\frac{1}{4}{\cal J}_{0}, {\cal J}_{2}=\frac{1}{4}{\cal J}_{0}, {\cal J}_{3}={\cal J}_{0}$.
Thus, the Hamiltonian of core can be written as
\begin{equation}\label{H2}
   \hat{H}_{\rm core} = \dfrac{1}{2{\cal J}_{0}} \left[4(\hat{R}^{2}_{1}+\hat{R}^{2}_{2})+\hat{R}^{2}_{3}\right].
\end{equation}
The intrinsic Hamiltonians $\hat{H}_{\rm p}$ and $\hat{H}_{n}$ describe the valance proton and
neutron outside the rotor. For a single-$j$ model, when pairing correlations are
neglected, $\hat{H}_{p}$ and $\hat{H}_{n}$ can be
given as
 \begin{eqnarray}\label{H3}
  \hat{H}_{\rm p(n)}&=&\pm \dfrac{1}{2}{C_0}\Big[(\hat{j}^{2}_{3}-\dfrac{j(j+1)}{3})\cos\gamma
   + \dfrac{1}{2\sqrt{3}}(\hat{j}^{2}_{+}+\hat{j}^{2}_{-})\sin\gamma \Big],
  \end{eqnarray}
where $C_{0}$ take values of $\dfrac{38.8(N+3/2)}{j(j+1)}A^{-1/3}\beta$~\cite{Zhang07,WangSY08}. For $\gamma=90^{\circ}$, it can be written as
  \begin{equation}\label{H4}
  \hat{H}_{\rm p(n)} = \pm \dfrac{1}{2\sqrt{3}}{C_0}(\hat{j}^{2}_{1}-\hat{j}^{2}_{2}).
\end{equation}

The total wave functions of the PRM Hamiltonian can be expanded in the strong coupling basis. Usually, the strong coupling basis
is expressed by \cite{Zhang07,Qi0901,Ikuko13,Ragnarsson88}
\begin{eqnarray}{}
  |IMK\varphi_{p}\varphi_{n}\rangle = \sqrt{\dfrac{1}{2}}\Big[|IMK\rangle |\varphi_{p}\varphi_{n}\rangle
              +(-1)^{I-K}|IM-K\rangle| \bar{\varphi}_{p}\bar{\varphi}_{n} \rangle \Big],
\end{eqnarray}
where $|IMK\rangle$ denotes the Wigner $D$ functions, $\varphi_{p}, \varphi_{n}$ and the time reversed states $\bar{\varphi}_{p}, \bar{\varphi}_{n}$ are the single-particle (or quasi-particle) eigenstates of intrinsic
Hamiltonian.

In the present paper, we adopt the basis as~\cite{PengJ03}
\begin{eqnarray}\label{kpkn}
  |IMKk_{ p}k_{n}\rangle= \sqrt{\dfrac{1}{2}}\Big[|IMK\rangle |k_{p}k_{n}\rangle
      +(-1)^{I-j_{p}-j_{n}}|IM-K\rangle|-k_{p} -k_{n}\rangle \Big],
  \end{eqnarray}
where $|k_{p}\rangle$ ($|k_{n}\rangle$) denotes the spherical harmonic oscillator state $|nljk\rangle$. For such basis of Eq.~(\ref{kpkn}), we can get the certain value of the third component of core angular momentum ($R_3=K-k_p-k_n$), which is necessary in the analysis for the operator of rotation by $90^{\circ}$ about 3-axis. Meantime, the operator of exchange valance proton and neutron could be dealt with by the exchange of the value of $k_p$
and $k_n$ of Eq.~(\ref{kpkn}). Thus it is easy to examine the symmetry under operator $\hat{A}$ for PRM wave functions by selecting this basis.

The probabilities of electromagnetic transition $B(M1)$ and $B(E2)$ can be obtained from the PRM wave functions with $M1$ and $E2$ operators~\cite{PengJ03,Zhang07,Koike04}.
For the $E2$ transitions, the corresponding operator is taken as
\begin{equation}\label{E2}
\hat{E}2=\sqrt{\dfrac{5}{16\pi}}\left[D^{2*}_{\mu0}\hat{Q}'_{20}+(D^{2*}_{\mu2}+D^{2*}_{\mu-2})\hat{Q}'_{22}\right],
\end{equation}
where $\hat{Q}'_{20}$ and $\hat{Q}'_{22}$ are the intrinsic quadrupole moments. For the $M1$ transitions, the corresponding operator is taken as
\begin{equation}\label{M1}
   (\hat{M}1)_{\mu}=\sqrt{\dfrac{3}{4\pi}}\dfrac{e\hbar}{2mc}\left[(g_{p}-g_{R})\hat{j}_{p\mu}+(g_{n}-g_{R})\hat{j}_{n\mu}\right]
\end{equation}
with
\begin{equation}\label{}
   \hat{j}_{\mu}=\left(\hat{j}_{0}=\hat{j}_{3},\hat{j}_{\pm1}=\dfrac{\mp(\hat{j}_{1}\pm i\hat{j}_{_{2}})}{\sqrt{2}}\right).
\end{equation}

In our calculations, the configurations $\pi g_{9/2} \otimes \nu g^{-1}_{9/2}$, $\pi h_{11/2} \otimes \nu h^{-1}_{11/2}$, $\pi i_{13/2} \otimes \nu i^{-1}_{13/2}$ for $A\sim$ 80, 130, 190 mass regions with deformation parameters $\beta = 0.22$, $\gamma = 90^{\circ}$ and moment of inertia ${\cal J}_{0} = 30MeV^{-1}\hbar^{2}$ are adopted. The empirical intrinsic quadrupole moment $Q_{0} = (3/\sqrt{5\pi})R^{2}_{0}Z\beta$ are adopted in the calculation of electromagnetic transitions. Using $g_{R} = Z/A$ and the empirical formula $g_{p(n)}=g_{l}+(g_{s}-g_{l})/(2l+1)$ with $g_{s}=0.6g^{free}_{s}$,  the $g$-factor for proton (neutron) occupied orbits $g_{9/2}, h_{11/2}, i_{13/2}$ would be $g_{p}(g_{n})-g_{R} \approx 0.82(-0.70), 0.77(-0.65), 0.76(-0.60)$, respectively.
Here, we take the approximate values $g_{p}(g_{n})-g_{R} = 0.7(-0.7)$ for all the occupying orbits with the intention of deducing the strictly forbidden $M1$ transitions in Eq.(10).

\section{Result and Discussion}

\begin{figure}[ht!]
\centering
\includegraphics[width=16cm]{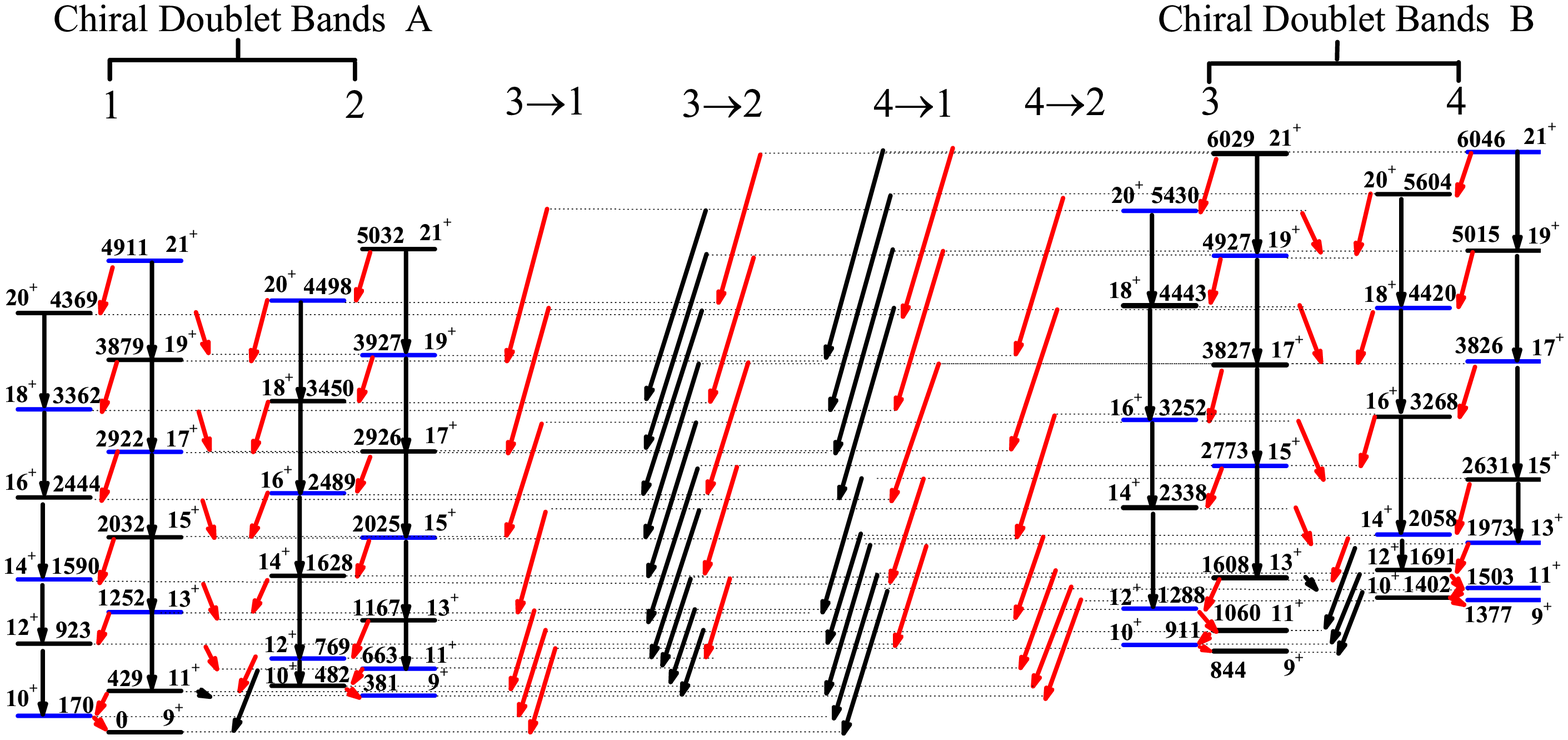}
\caption{Calculated level scheme for two pairs of chiral doublet bands based on the configuration $\pi h_{11/2}\otimes \nu h_{11/2}^{-1}$ coupled with  $\gamma =90^{\circ}$ rotor. Black and blue solid lines represent the energy levels with $A = 1$ and $-1$. The parity quantum numbers, the angular momentum quantum numbers and excited energies are listed above the levels. Red arrows represent $M1(I\rightarrow I-1)$ transitions and black arrows represent $E2(I\rightarrow I-2)$ transitions.}
\label{fig1}
\end{figure}

\begin{figure}[ht!]
\centering
 \includegraphics[width=16cm]{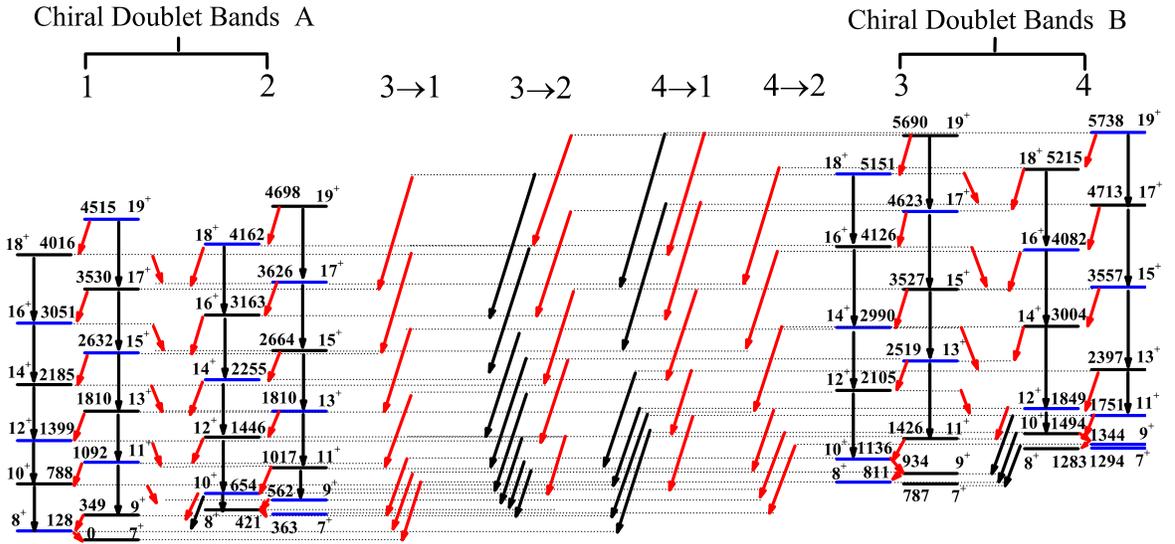}
\caption{Same as Fig.\ref{fig1} but for the configuration $\pi g_{9/2}\otimes \nu g_{9/2}^{-1}$.}
\label{fig2}
\end{figure}

\begin{figure}[ht!]
\centering
  \includegraphics[width=16cm]{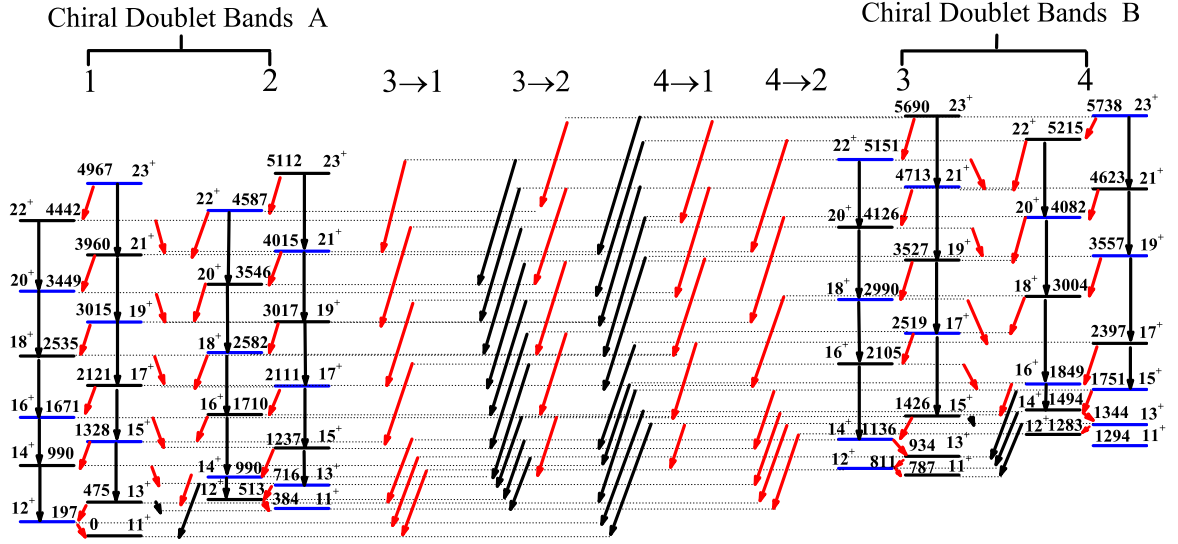}
\caption{Same as Fig.\ref{fig1} but for the configuration  $\pi i_{13/2}\otimes \nu i_{13/2}^{-1}$.}
\label{fig3}
\end{figure}

The calculated  level scheme for two pairs of chiral doublet bands based on the configurations $\pi h_{11/2}\otimes \nu h_{11/2}^{-1}$, $\pi g_{9/2} \otimes \nu g^{-1}_{9/2}$ and $\pi i_{13/2} \otimes \nu i^{-1}_{13/2}$ coupled with $\gamma$ =90$^\circ$ rotor are shown in Figs.~\ref{fig1},~\ref{fig2} and~\ref{fig3}.
The parity quantum numbers, the angular momentum quantum numbers  and excited energies are listed above the energy levels. Red arrows represent $M1(I\rightarrow I-1)$ transitions and black arrows represent $E2(I\rightarrow I-2)$ transitions. The bands are organized based on in-band $B(E2;I\rightarrow I-2)$ values over the degenerate spin range, namely  $I\geq 10, 12, 14\hbar$ for 80, 130, 190 mass regions, respectively. These bands are labeled as 1,~2,~3,~4. Bands 1 $\&$ 2 form the lowest chiral doublet bands
A, while Bands 3 $\&$ 4 form the excited chiral doublet bands B.

Taking the 130 mass region as an example, the properties of electromagnetic transitions in  M$\chi$D are discussed. The corresponding $B(M1;I\rightarrow I-1)$ and $B(E2;I\rightarrow I-2)$ values are presented in Tables.~\ref{tab1} and~\ref{tab2}.
As shown in Fig.~\ref{fig1}, the bands are organized based on $B(E2)$ values so that the in-band $E2$ transitions are always allowed. The interband $E2$ transitions are allowed from the states of the Band 3 decaying to those of the Band 2, or Band 4 to 1. For the in-band $M1$ transitions, the
same odd-even spin staggering is clearly seen in the four bands, in which transitions from odd spin to even spin states are allowed. For the interband $M1$ transitions, transitions from Band 1 to 2 and Band 2 to 1 are allowed for even spin states decaying to odd spin states, with the same behavior exhibited for Band 3 and 4. Interband $M1$ transitions between Chiral Bands A and Chiral Bands B are allowed from the states of Band 3 decaying alternatively to those of the Band 1 or 2, with the same behavior exhibited for Band 4.

The above selection rules of electromagnetic transitions associated with odd and even spin are summarized in Table.~\ref{tab3}.
For the 80 and 190 mass regions, the similar proprieties, especially the same selection rule associated with odd and even spin as the case of the 130 mass region, are obtained from the model calculations.

Besides the selection rule for M$\chi$D, the quantitative relations of electromagnetic transitions probabilities are also obtained from the Tables.~\ref{tab1} and~\ref{tab2}. The $B(M1)$ and $B(E2)$ values in the excited chiral doublet bands have the same order of magnitude as those in lowest chiral doublet bands. However, the $B(M1)$ and $B(E2)$ values linked the excited to the lowest chiral doublet bands are two orders of magnitude smaller than those in the lowest(or excited) chiral doublet bands. The selection rule and the quantitative relations of electromagnetic transitions probabilities would be helpful for confirming the existence of the M$\chi$D in the real nuclei.

As discussed in Ref.~\cite{Koike04} for the ideal case with one-particle one-hole plus a $\gamma=90^{\circ}$ rotor,
the chiral operator $\chi=\hat{T}\hat{R}_2(\pi)$ can be replaced by the operator $\hat{A}$  defined as
\begin{equation}\label{}
   \hat{A}=e^{i\dfrac{\pi}{2}\hat{R}_{3}}\cdot \hat{C},
\end{equation}
where the operator $e^{i\dfrac{\pi}{2}\hat{R}_{3}}$ denotes core rotation by $90^{\circ}$ about the 3-axis and the operator $\hat{C}$ denotes the exchange of the valance proton and neutron. It is obvious that the PRM Hamiltonian described by Eqs.~(\ref{H2}) and (\ref{H4}) is symmetric under the operator $\hat{A}$. Then according to Quantum Mechanics~\cite{Sakurai07}, the wave function might have such a symmetry, or bring spontaneous symmetry breaking. Thus, it is necessary to examine the symmetry
of wave function as the first step.

Taking the state of $I= 17\hbar$ of Band 1 for the 130 mass region for example, the corresponding calculated wave functions of the PRM are listed in Table.~\ref{tab4}. $|k_{p},k_{n},K\rangle$ denotes the basis in Eq.~(\ref{kpkn}). $k_{p}, k_{n}, K$ and $ R_{3} (R_{3}=K-k_{p}-k_{n})$ refer to the third component of angular momentum for the valance proton, valance neutron, nucleus and core, respectively. $C^{IK}_{k_{p},k_{n}}$ refers to the expansion coefficient of the basis. The quantum number
$R_{3}$  takes only even integer values due to $D_{2}$ symmetry. $C$ takes 1 and $-1$ due to the symmetry and antisymmetry of the intrinsic proton-neutron wave function under the exchange of proton and neutron, respectively. The $A$ values  can be fixed by ($n=0,1,2,3..$)
\begin{enumerate}
  \item $R_3= \pm4n$, $C=1$, or $R_3= \pm4n +2$, $C=-1$, resulting in $A=1$,
  \item $R_3= \pm4n$, $C=-1$, or $R_3= \pm4n +2$, $C=1$, resulting in $A=-1$.
\end{enumerate}
The components of the wave function are divided into two groups: one with $C=1$, such as $(0.111|2.5,3.5, 16\rangle+0.111 |3.5,2.5,16\rangle)$, and the other with $C=-1$,  such as $(-0.107|-0.5,1.5,13\rangle+0.107|1.5,-0.5,13\rangle)$. Together with the values of corresponding $R_3$ in the different components of the wave function, $A=-1$ can be obtained for $I= 17\hbar$ of Band 1. The wave function is expanded in a 1260 $\left( \sim(2j_p+1)(2j_n+1)(2I+1)/4\right )$ dimensional basis for such a state, in which all components meet the symmetry with $A=-1$.
The calculated PRM wave functions of M$\chi$D in the three mass regions are systematically examined, and are found to
always have eigenvalues $A$ = 1 or $-1$.

Therefore, levels with $A$ = 1 and $-1$ are expressed by black and blue solid lines in Figs.~\ref{fig1},~2,~3. We found that $M1$ and $E2$ transitions are allowed between levels with different values of $A$, but forbidden between the levels with the same values of $A$. Such a selection rule was examined for the lowest chiral doublet bands in the 130 mass region~\cite{Koike04}. The present results show that the selection rule is not only suitable for the lowest chiral doublet bands but also suitable for the excited chiral doublet bands, not only in the 130 mass region but also in the 80 and 190 mass regions with symmetric proton-neutron configurations.

The quantum number $A$ and the resulting selection rule are attributed to the symmetry under operator $\hat{A}$ for the Hamiltonian with the present ideal case. For the Hamiltonian with the asymmetric configuration or $\gamma$ deviating from 90$^{\circ}$, it is obvious that $A$ is no longer a good quantum number and the resulting selection rule could not be obtained. However, by analyzing the previous results~\cite{Chen16, Qi0902} and our systematical calculations, it is found that the selection rule associated with odd-even spin still maintains for the lowest chrial doublet bands when $\gamma$ is close to 90$^{\circ}$, while  is difficult to be kept for the excited chiral doublet bands.
So far, the only case of M$\chi$D with identical intrinsic configuration was observed in odd-$A$ nucleus $^{103}$Rh with configuration $\pi g^{-1}_{9/2}\otimes \nu h_{11/2}g_{7/2}$~\cite{Kuti14}. The $B(M1)/B(E2)$ ratios of M$\chi$D in  $^{103}$Rh  were extracted and exhibited weak staggering~\cite{Kuti14}, which show the deviations  from the present ideal cases. The deviations might be attributed to the asymmetric configuration and $\gamma$ deviated far from 90$^{\circ}$.

The reason for the selection rule for electromagnetic transition in terms of ${A}$ values has been discussed in Ref.~\cite{Koike04}. Here, some more detailed explanation is discussed.
For the $E2$ transition operator in Eq.~(\ref{E2}), the $B(E2)$ values can be obtained as following, if only the core contributions are considered~\cite{PengJ03,Koike04}

\begin{eqnarray}\label{}
B(E2,I\rightarrow I') &=&Q^{2}_{0}\dfrac{5}{16\pi}|\sum^{k_{p}k_{n}}_{K,K'}C^{IK}_{k_{p}k_{n}}C^{I'K'}_{k_{p}k_{n}}[\cos\gamma\langle IK20|I'K'\rangle \nonumber\\
&-&\dfrac{\sin\gamma}{\sqrt{2}}(\langle IK22|I'K'\rangle+\langle IK2-2|I'K'\rangle)]|^{2}.
\end{eqnarray}

To get the non-zero $ E2$  matrix elements, the wave function of the valence proton and neutron in the initial state should be the same as the final state. This means initial state and final state must have the same symmetry under the exchange of valance proton and neutron ($\Delta C = 0$). Furthermore, $\gamma = 90^{\circ}$ means that $\cos\gamma\langle IK20|I'K'\rangle=0$. Therefore, only the $E2$ matrix elements with $\Delta C = 0$ and $\Delta R_{3}=K'-K = \pm 2$ are non-zero.
The non-zero $E2$ matrix elements connect states of $R_3= \pm4n$, $C=1$ ($A=1$) with $R_3= \pm4n +2$, $C=1$ ($A=-1$), or connect states of $R_3= \pm4n$, $C=-1(A=-1)$ with  $R_3= \pm4n +2$, $C=-1$ ($A=1$). Thus  the $E2$ transitions are allowed between levels with different values of $A$, while forbidden with the same values of $A$.

For the $M1$ transition operator shown in Eq.~(\ref{M1}), $\hat{M}1 \propto \hat{j}_{p(\mu)}-\hat{j}_{n(\mu)} (\mu=0, \pm1 )$ when we take $g_{p}-g_{R}= -(g_{n}-g_{R})$ for orbits $g_{9/2}, h_{11/2}, i_{13/2}$. According to the following relationships

\begin{eqnarray}
&&(\hat{j}_{p(0)}-\hat{j}_{n(0)})(|k_{p} k_{n}\rangle+|k_{n} k_{p}\rangle)\nonumber\\
&=&(k_{p}-k_{n})(|k_{p} k_{n}\rangle-|k_{n} k_{p}\rangle),\nonumber\\
&&(\hat{j}_{p(+1)}-\hat{j}_{n(+1)})(|k_{p} k_{n}\rangle+|k_{n} k_{p}\rangle)\nonumber\\
&=&\sqrt{\dfrac{j(j+1)-k_{p}(k_{p}+1)}{2}}(|k_{n}, k_{p}+1\rangle-|k_{p}+1, k_{n}\rangle)\nonumber\\
&+&\sqrt{\dfrac{j(j+1)-k_{n}(k_{n}+1)}{2}}(|k_{p},k_{n}+1\rangle-|k_{n}+1, k_{p}\rangle),\nonumber\\
&&(\hat{j}_{p(-1)}-\hat{j}_{n(-1)})(|k_{p} k_{n}\rangle+|k_{n} k_{p}\rangle)\nonumber\\
&=&\sqrt{\dfrac{j(j+1)-k_{p}(k_{p}-1)}{2}}(|k_{p}-1, k_{n}\rangle-|k_{n}, k_{p}-1\emph{}\rangle)\nonumber\\
&+&\sqrt{\dfrac{j(j+1)-k_{n}(k_{n}-1)}{2}}(|k_{n}-1, k_{p}\rangle-|k_{p},k_{n}-1\rangle),\nonumber\\
\end{eqnarray}
the component of the wave function with $C = 1$, acted on by $\hat{M}1$ transition operators, will change to ones with $C = -1$, and vice versa. Therefore, $M1$ matrix elements between the initial and final states
with the same $C$ values will be zero exactly. Because the $M1$ operator only connects
components with $\Delta R_{3}= 0$, only the $M1$ transitions
between states with different $A$ values (same $R_3$ and opposite $C$ values ) are allowed, while those
between states with the same $A$ values are forbidden.  In the real cases, $g_{p}-g_{R}$ is slightly different from $-(g_{n}-g_{R})$, so $M1$ transition probabilities
between states with the same $A$ values are about an
order of magnitude smaller than those with opposite $A$.

The present selection rule is applicable to M$\chi$D with the identical symmetric configuration in odd-odd nuclei.  The observation of the M$\chi$D with the identical symmetric configuration in the odd-odd nuclei is expected in the future experiment to examine the present selection rule.

\section{Summary}

M$\chi$D based on $\gamma$=90$^{\circ}$ triaxial rotor coupled with identical symmetric proton-neutron configurations are studied by adopting the particle rotor model, in which the configurations are $\pi g_{9/2} \otimes \nu g^{-1}_{9/2}$, $\pi h_{11/2} \otimes \nu h^{-1}_{11/2}$ and $\pi i_{13/2} \otimes \nu i^{-1}_{13/2}$ for the $80, 130$ and $190$ mass regions, respectively. The calculated wave functions of  M$\chi$D are quantitatively analyzed for the first time and found to meet strict symmetry under the operator $\hat{A}$. The selection rule for the excited chiral bands and for the different mass regions are examined by the numerical results.
The selection rule for the electromagnetic transitions, namely $M1$ and $E2$ transitions are allowed between the levels with different values of $A$  while forbidden (or much weaker for $M1$) between the levels with the same values of $A$, holds true for the M$\chi$D with symmetric proton-neutron configurations. The present results might be helpful to identify the M$\chi$D in experiment.

\section*{Acknowledgements}
This work is partly supported by National Natural Science
Foundation of China (11675094, 11622540, 11545011, 11405096, 11461141001, U1432119), the Shandong Natural Science Foundation (ZR2014AQ012), and the Young Scholars Program of Shandong University,
Weihai(2015WHWLJH01).

\clearpage


\clearpage

\begin{table}[pt]
\caption{\label{tab1}Calculated $B(M1;I\rightarrow I-1)(\mu_{N}^{2})$ values for  two pairs of chiral doublet bands based on the configuration $\pi h_{11/2}\otimes \nu h_{11/2}^{-1}$ coupled with  $\gamma =90^{\circ}$ rotor.}
\begin{tabular*}{117mm}{@{}c|cccccccccccc@{}}
\toprule
\hline
                            & \multicolumn{11}{c}{ $I(\hbar)$} & \\
                                  \cline{2-13}
                                        &  10 &  11   & 12    & 13    &14     & 15    & 16    & 17    & 18    & 19    & 20    & 21  \\
                                        \hline
                          ~1$\rightarrow$1~ & 3.1 & 2.9 &  0 & 2.7 & 0 & 2.4 & 0 & 2.4 & 0 & 2.2 & 0 & 2.1  \\
                          1$\rightarrow$2   & 0 & 0& 2.9 & 0& 2.3 &0 & 2.5 & 0 & 2.5& 0 & 2.4 & 0     \\\hline
                          2$\rightarrow$1 &  0 & 0 & 2.7 & 0& 2.6 & 0& 2.3 & 0 & 2.2 & 0& 1.9 & 0   \\
                          2$\rightarrow$2 &  3.3 & 3.1 & 0 & 2.5 & 0 & 2.5 & 0 & 2.4 & 0& 2.3 & 0 & 2.0  \\\hline
                          3$\rightarrow$1 &  0.002 & 0.01 & 0.03 & 0 & 0.04 & 0 & 0.05 & 0 & 0.03& 0 & 0.03 & 0    \\
                          3$\rightarrow$2 &  0& 0 & 0& 0.10 & 0 & 0.06 & 0 & 0.05 & 0 & 0.05 & 0 & 0.09   \\
                          3$\rightarrow$3 &  3.3 & 3.1 & 2.9 & 2.6 & 0 & 2.3 & 0 & 1.7 & 0 & 1.9 & 0 & 1.2      \\
                          3$\rightarrow$4 &  0 & 0 & 0 &0 & 2.5 & 0 & 1.9& 0 & 1.5 & 0 & 1.3 & 0      \\\hline
                          4$\rightarrow$1 &  0 & 0 & 0& 0.02 & 0 & 0.09 & 0& 0.02 & 0 & 0.0002 & 0 & 0.003       \\
                          4$\rightarrow$2 &  0.001 & 0.01 & 0.01 & 0 & 0.16 & 0 & 0.05 & 0 & 0.04 & 0 & 0.02 & 0    \\
                          4$\rightarrow$3 & 0 & 0 & 0 & 0 & 2.2 & 0 & 2.1 & 0 & 1.9 & 0 & 1.6 & 0    \\
                          4$\rightarrow$4 &  3.3 & 3.1 & 3.0 & 2.7 & 0 & 2.0 & 0 & 2.0 & 0 & 0.02 & 0 & 0.09           \\
                                   \bottomrule
                                   \hline
\end{tabular*}%
\end{table}

\begin{table}[pt]
\caption{\label{tab2}Calculated $B(E2;I\rightarrow I-2)$(e$^{2}$b$^{2}$) values for  two pairs of chiral doublet bands based on the configuration $\pi h_{11/2}\otimes \nu h_{11/2}^{-1}$ coupled with  $\gamma =90^{\circ}$ rotor.}
\footnotesize
\begin{tabular*}{117mm}{@{}c|cccccccccccc@{}}
\toprule
\hline
                              & \multicolumn{11}{c}{ I($\hbar$)} &   \\
                                  \cline{2-13}
                                                 & 11 & 12    & 13    &14     & 15    & 16    & 17    & 18    & 19    & 20    & 21 &       \\ \hline
                                  1$\rightarrow$1&  0 & 0.12 & 0.12 & 0.18 & 0.23 & 0.25 & 0.30 & 0.35 & 0.37 & 0.40   & 0.42 &     \\
                                  1$\rightarrow$2 & 0.04 &  0 & 0 & 0 & 0 & 0 & 0 & 0& 0 & 0   &   0   &    \\\hline
                                  2$\rightarrow$1 & 0.12 & 0 & 0 & 0 & 0 & 0 & 0 & 0 & 0 & 0 & 0&             \\
                                  2$\rightarrow$2 & 0 & 0.06 & 0.11 & 0.14 & 0.19 & 0.27 & 0.31 & 0.33 & 0.37 & 0.39 & 0.42 &           \\\hline
                                  3$\rightarrow$1 & 0 & 0 & 0 & 0 & 0 & 0 & 0 & 0 & 0 & 0 & 0 & \\
                                  3$\rightarrow$2 & 0.17 & 0.15 & 0.10 & 0.001 & 0.001 & 0.004 & 0.001 & 0.001 & 0.0002 & 0.004 &  0 & \\
                                  3$\rightarrow$3 & 0 & 0 & 0 & 0.16 & 0.13 & 0.16 & 0.22 & 0.20 & 0.25 & 0.10 & 0.25 & \\
                                  3$\rightarrow$4 & 0.03 & 0.04 & 0.05  & 0 & 0 & 0 & 0& 0 & 0 & 0 & 0 &  \\ \hline
                                  4$\rightarrow$1 & 0.0002 & 0.001 & 0.001 & 0.05 & 0.01 & 0 & 0 & 0.0003 & 0.01 & 0.0003 & 0.01 & \\
                                  4$\rightarrow$2 & 0 & 0 & 0 & 0 & 0 & 0 & 0 & 0 & 0& 0 & 0 & \\
                                  4$\rightarrow$3 & 0.23 & 0.21 & 0.18  & 0 & 0 & 0 & 0 & 0 & 0 & 0 & 0 &  \\
                                  4$\rightarrow$4 & 0 & 0 & 0& 0.07 & 0.10 & 0.12 & 0.14 & 0.24 & 0.01 & 0.32 & 0.41& \\
                                            \bottomrule
                                            \hline
\end{tabular*}%
\end{table}

\begin{table}[pt]
\caption{\label{tab3}Selection rule of $M1$ and $E2$ transitions associated with odd-even initial spins.}
\footnotesize
\begin{tabular*}{82mm}{c|c|c|c|c}
\toprule
\hline
       & \multicolumn{2}{c}{ $M1(I\rightarrow I-1)$} & \multicolumn{2}{c}{ $E2(I\rightarrow I-2)$}  \\
 \cline{1-5}
   $I$    &  odd spin  &  even spin & odd spin &   even spin \\
\hline
     in-band &   \multirow{3}{*}{allowed} &\multirow{3}{*}{forbidden}  &  \multirow{3}{*}{allowed} &\multirow{3}{*}{allowed} \\
Band 3$\rightarrow$  2    &  &                 & & \\
     Band 4$\rightarrow$  1  & &  & &  \\ \hline
      Band 1$\leftrightarrow$  2 & \multirow{4}{*}{forbidden}   &\multirow{4}{*}{allowed}  & \multirow{4}{*}{forbidden}  &\multirow{4}{*}{forbidden}\\
                 Band 3$\leftrightarrow$  4  & &  & &  \\
                  Band 3$\rightarrow$  1  & &   & &   \\
                   Band 4$\rightarrow$  2  & &   & &\\
 \bottomrule
 \hline
\end{tabular*}
\end{table}

\begin{table}[pt]
\caption{\label{tab4}Calculated wave functions at $I= 17\hbar$ of Bands 1-4 based on the configuration $\pi h_{11/2}\otimes \nu h_{11/2}^{-1}$ coupled with  $\gamma =90^{\circ}$ rotor. $|k_{p},k_{n},K\rangle$ is the basis of wave function. $C^{IK}_{k_{p},k_{n}}$ refers to expansion coefficient of basis. $R_{3}$ refers to third component of core angular momentum.  $C$ is quantum number for the operator of exchange proton and neutron, and $A=exp^{i\frac{\pi}{2}R_{3}}\cdot C$.}
\begin{tabular*}{120mm}{cccccccccc|ccccccccccc}
\toprule
\hline
   \multicolumn{9}{c}{ Band 1} & &\multicolumn{9}{c}{ Band 2}\\
    \hline
  && $C^{IK}_{k_{p},k_{n}}|k_{p},k_{n},K\rangle$& &  $R_{3}$ & & $C$ & &$A$    &      &  $C^{IK}_{k_{p},k_{n}}|k_{p},k_{n},K\rangle$& &  $R_{3}$ & & $C$ & &  $A$ & &   \\
\hline
   & & $-0.107|-0.5,1.5,13\rangle$  & & 12& & \multirow{2}{*}{-1} && \multirow{2}{*}{-1} & &$0.106|1.5,2.5,12\rangle$  & & 8&&\multirow{2}{*}{1} && \multirow{2}{*}{1} \\
   & & $ 0.107|1.5,-0.5,13\rangle$ & &12& &   & &                                        & & $0.106|2.5,1.5,12\rangle$ & &8& &   &&     \\
   & & $0.145|2.5,4.5,15\rangle$  & & 8& & \multirow{2}{*}{-1} && \multirow{2}{*}{-1} & &  $-0.111|2.5,3.5,16\rangle$  & & 10& & \multirow{2}{*}{-1} && \multirow{2}{*}{1} \\
   & & $-0.145|4.5,2.5,15\rangle$ & &8& &   & &                                         & &$ 0.111|3.5,2.5,16\rangle$ & &10& &   & &           \\
   & & $0.111|2.5,3.5, 16\rangle$  & &10 & & \multirow{2}{*}{1} && \multirow{2}{*}{-1}   & &$-0.100|2.5,4.5,17\rangle$  & & 10 & & \multirow{2}{*}{-1} && \multirow{2}{*}{1}\\
   & & $0.111 |3.5,2.5,16\rangle$ & &10& &   & &                                         & &$0.100 |4.5,2.5,17 \rangle$ & &10& &   & & \\
    & & ... & &...& &...   & &    ...         &&  ... & &..& & ...  & & ... & &     \\
      & & $-0.004|-4.5,5.5,15\rangle$& & 14& & \multirow{2}{*}{1} && \multirow{2}{*}{-1} & &    $-0.005|-3.5,4.5,15\rangle$& & 14&&\multirow{2}{*}{-1} && \multirow{2}{*}{1} \\
   & & $-0.004|5.5,-4.5,15 \rangle$ & &14& &   & &                                     & &  $ 0.005 |4.5,-3.5,15\rangle$ & &14& &   & &  \\
   & & $-0.009|-3.5,5.5,16\rangle$  & & 14& & \multirow{2}{*}{1} && \multirow{2}{*}{-1} & &  $0.009|-3.5,5.5,16\rangle$  & & 14& & \multirow{2}{*}{-1} && \multirow{2}{*}{1} \\
   & & $-0.009|5.5,-3.5,16\rangle$ & &14& &   & &                                         & &$ -0.009|5.5,-3.5,16\rangle$ & &14& &   & &           \\
   & & ... & &...& &...   & &    ...         &&  ... & &..& & ...  & & ... & &     \\
\hline
\multicolumn{9}{c}{ Band 3} & &\multicolumn{9}{c}{ Band 4}\\
\hline
   & &  $ 0.107|0.5,3.5,10\rangle$  & & 6& & \multirow{2}{*}{-1} && \multirow{2}{*}{1}    &&$0.122|1.5,2.5,10\rangle$  & & 6& &\multirow{2}{*}{1} &&\multirow{2}{*}{-1}\\
   & &  $ -0.107|3.5,0.5,10\rangle$  & & 6& &  & &                                        & &$0.122|2.5,1.5,10\rangle$  & & 6& &\\
    & &  $-0.103|0.5,3.5,16\rangle$  & & 12 & & \multirow{2}{*}{1} && \multirow{2}{*}{1}  && $-0.112|0.5,2.5,11\rangle$  & & 8 & & \multirow{2}{*}{-1}&&\multirow{2}{*}{-1}\\
  & &   $-0.103|3.5,0.5,16\rangle$ & &12& &   & &                                  & &       $0.112|2.5,0.5,11\rangle$ & &8&&   & & \\
    & &  $ 0.121|2.5,4.5,17\rangle$  & & 10 & & \multirow{2}{*}{-1} && \multirow{2}{*}{1}  &&$ 0.104|3.5,5.5,17\rangle$  & & 8 & & \multirow{2}{*}{-1} &&\multirow{2}{*}{-1}\\
    & &   $-0.121|4.5,2.5,17\rangle$ & &10& &   & &                                    & &   $-0.104|5.5,3.5,17\rangle$ & &8&&   & & \\
     & & ... & &...& &...   & &    ...         &&  ... & &..& & ...  & & ... & &     \\
        & & $-0.005|-3.5,4.5,15\rangle$& & 14& & \multirow{2}{*}{-1} && \multirow{2}{*}{1} & &    $-0.014|2.5,3.5,14\rangle$& & 8&&\multirow{2}{*}{-1} && \multirow{2}{*}{-1} \\
   & & $0.005|4.5,-3.5,15 \rangle$ & &14& &   & &                                     & &  $ 0.014 |3.5,2.5,14\rangle$ & &8& &   & &  \\
   & & $0.005|-4.5,5.5, 17\rangle$  & &16 & & \multirow{2}{*}{1} && \multirow{2}{*}{1}   & &$0.003|-4.5,5.5,17\rangle$  & & 16 & & \multirow{2}{*}{-1} && \multirow{2}{*}{-1}\\
   & & $0.005|5.5,-4.5,17\rangle$ & &16& &   & &                                         & &$-0.003 |5.5,-4.5,17 \rangle$ & &16& &   & & \\
   & & ... & &...& &...   & &    ...         &&  ... & &..& & ...  & & ... & &     \\
            \bottomrule
            \hline
\end{tabular*}%
\end{table}


\begin{thebibliography}{100}
\bibitem{FM97} S.~Frauendorf and J.~Meng, Nucl. Phys. A, {\bf 617}: 31 (1997).
\bibitem{Starosta01} K. Starosta  et al., Phys. Rev. Lett., {\bf 86}: 971 (2001).
\bibitem{Koike01} T. Koike, K. Starosta, C. J. Chiara, D. B. Fossan, and D. R. LaFosse, Phys. Rev. C, {\bf63}: 061304(R) (2001).
\bibitem{Bark01} R. A. Bark, A.M. Baxter, A.P. Byrne, G.D. Dracoulis, T. Kib\'{e}di, T.R. McGoram, S.M. Mullins, Nucl. Phys. A, {\bf 691}: 577 (2001).
\bibitem{Hecht01} A. A. Hecht et al., Phys. Rev. C, {\bf63}: 051302(R) (2001).
\bibitem{Hartley01} D. J. Hartley et al., Phys. Rev. C, {\bf64}: 031304(R) (2001).
\bibitem{Mergel02} E. Mergel et al., Eur. Phys. J. A, {\bf15}: 417 (2002).
\bibitem{Koike03} T. Koike, K. Starosta, C. J. Chiara, D. B. Fossan, and D. R. LaFosse, Phys. Rev. C, {\bf67}: 044319 (2003).
\bibitem{Zhu03} S. Zhu et al., Phys. Rev. Lett., {\bf91}: 132501 (2003).
\bibitem{Vaman04} C. Vaman, D. B. Fossan, T. Koike, K. Starosta, I. Y. Lee, and A. O. Macchiavelli, Phys. Rev. Lett., {\bf92}: 032501 (2004).
\bibitem{Joshi04} P. Joshi et al., Phys. Lett. B, {\bf595}: 135 (2004).
\bibitem{Timar04} J. Tim\'{a}r et al., Phys. Lett. B, {\bf 598}: 178 (2004).
\bibitem{Alcantara04} J. A. Alc\'{a}ntara-N\'{u}\~{n}ez et al.,  Phys. Rev. C, {\bf69}: 024317 (2004).
\bibitem{WangSY06a} S. Y. Wang, Y. Z. Liu, T. Komatsubara, Y. J. Ma, and Y. H. Zhang, Phys. Rev. C, {\bf74}: 017302 (2006).
\bibitem{Grodner06} E. Grodner et al., Phys. Rev. Lett., {\bf97}: 172501 (2006).
\bibitem{Tonev06} D. Tonev et al., Phys. Rev. Lett., {\bf96}: 052501 (2006).
\bibitem{Timar06} J. Tim\'{a}r, C. Vaman, K. Starosta, D. B. Fossan, T. Koike, D. Sohler, I. Y. Lee, and A. O. Macchiavelli, Phys. Rev. C, {\bf73}: 011301(R) (2006).
\bibitem{Joshi07} P. Joshi, M. P. Carpenter, D. B. Fossan, T. Koike, E. S. Paul, G. Rainovski, K. Starosta, C. Vaman, and R. Wadsworth, Phys. Rev. Lett., {\bf98}: 102501 (2007).
\bibitem{Lawrie08} E. A. Lawrie et al., Phys. Rev. C, {\bf78}: 021305(R) (2008).
\bibitem{Suzuki08} T. Suzuki et al., Phys. Rev. C, {\bf78}: 031302(R) (2008).
\bibitem{Wang11} S. Y. Wang et al., Phys. Lett. B, {\bf703}: 40 (2011).
\bibitem{Koike05} T. Koike, K. Starosta, I. Hamamoto, D. B. Fossan and C. Vaman, AIP Conf. Proc., {\bf764}: 87 (2005).
\bibitem{Masiteng13} P.L. Masiteng et al., Phys. Lett. B, {\bf719}: 83 (2013).
\bibitem{Liu16} C. Liu et al., Phys. Rev. Lett., {\bf116}: 112501(2016).
\bibitem{Dimitrov00PRL} V. I. Dimitrov, S. Frauendorf, and F. D\"{o}nau, Phys. Rev. Lett., {\bf84}: 5732 (2000).
\bibitem{Olbratowski04} P. Olbratowski, J. Dobaczewski, J. Dudek, and W. Pl\'{o}iennik, Phys. Rev. Lett., {\bf 93}: 052501 (2004).
\bibitem{Mukhopadhyay07} S. Mukhopadhyay et al., Phys. Rev. Lett., {\bf99}: 172501 (2007).
\bibitem{PengJ03} J. Peng, J. Meng, and S. Q. Zhang, Phys. Rev. C, {\bf68}: 044324 (2003).
\bibitem{Zhang07} S. Q. Zhang, B. Qi, S. Y. Wang, and J. Meng, Phys. Rev. C, {\bf75}: 044307 (2007).
\bibitem{Wangsy07} Wang Shou-Yu, ZHANG Shuang-Quan, QI Bin, MENG Jie, Chin. Phys. Lett., {\bf24}: 664(2007).
\bibitem{Meng08} J. MENG£¬B. QI£¬S. Q. ZHANG£¬S. Y. WANG, Mod. Phys. Lett. A, {\bf23}: 2560 (2008).
\bibitem{Wang08} S. Y. Wang, S. Q. Zhang, B. Qi, and J. Meng, Chin. Phys. C, {\bf32}: 138 (2008).
\bibitem{WangSY08} S. Y. Wang, S. Q. Zhang, B. Qi, J. Peng, J. M. Yao, and J. Meng, Phys. Rev. C, {\bf77}: 034314 (2008).
\bibitem{Wang09} S. Y. Wang, S. Q. Zhang, B. Qi, and J. Meng, Chin. Phys. C, {\bf33}: 37 (2009).
\bibitem{Qi0901} B. Qi, S. Q. Zhang, J. Meng, and S. Frauendorf, Phys. Lett. B, {\bf675}: 175 (2009).
\bibitem{Wangsy10}S. Y. Wang, B. Qi, and D. P. Sun, Phys. Rev. C, {\bf82}: 027303 (2010).
\bibitem{Liul13} L. LIU, S. Y. WANG, B.QI and C. LIU, Int. J. Mod. Phys. E, {\bf22}: 1350060 (2013).
\bibitem{Qi12} QI Bin, ZHANG Pan, ZHANG Jing et al., Chin. Phys. C, {\bf36}: 10 (2012).
\bibitem{Tonev07} D. Tonev et al., Phys. Rev. C, {\bf76}: 044313 (2007).
\bibitem{Bhat14} G. H. Bhat et al., Phys. Lett. B, {\bf738}: 218 (2014),
\bibitem{MengJ06} J. Meng, J. Peng, S. Q. Zhang, and S. G. Zhou, Phys. Rev. C, {\bf73}: 037303 (2006).
\bibitem{Ayangeakaa13} A. D. Ayangeakaa et al., Phys. Rev. Lett., {\bf110}: 172504 (2013).
\bibitem{Qi13} B. Qi, H. Jia, N. B. Zhang, C. Liu, and S. Y. Wang, Phys. Rev. C, {\bf88}: 027302 (2013).
\bibitem{Droste09} Ch. Droste, S. G. Rohozinski, K. Starosta, L. Pr$\acute{o}$chniak, and
E. Grodner, Eur. Phys. J. A, {\bf42}: 79 (2009).
\bibitem{Qb10} Q. B. Chen and J. M. Yao, S. Q. Zhang, B. Qi, Phys. Rev. C, {\bf82}: 067302 (2010).
\bibitem{Ikuko13} I. Hamamoto, Phys. Rev. C, {\bf88}: 024327 (2013).
\bibitem{Chen16} Hao Zhang, QiBo CHEN, Chin. Phys. C, {\bf{40}}: 024102 (2016).
\bibitem{Kuti14} I. Kuti et al., Phys. Rev. Lett., {\bf113}: 032501 (2014).
\bibitem{Peng08} J. Peng, J. Meng, P. Ring, and S. Q. Zhang, Phys. Rev. C, {\bf78}: 024313 (2008).
\bibitem{Zhao11} P.W. Zhao, S. Q. Zhang, J. Peng, H. Z. Liang, P. Ring, and J. Meng, Phys. Lett. B, {\bf699}: 181 (2011).
\bibitem{Zhao111} P.W. Zhao, J. Peng, H. Z. Liang, P. Ring, and J. Meng, Phys. Rev. Lett., {\bf107}: 122501 (2011).
\bibitem{Zhao12} P.W. Zhao, J. Peng, H. Z. Liang, P. Ring, and J. Meng, Phys. Rev. C, {\bf85}: 054310 (2012).
\bibitem{Meng13} J. Meng, J. Peng, S. Q. Zhang, and P.W. Zhao, Front. Phys., {\bf8}: 55 (2013).
\bibitem{Peng15} J.Peng and P. W. Zhao, Phys.Rev. C, {\bf91}: 044329 (2015).


\bibitem{Koike04} T. Koike, K. Starosta, and I. Hamamoto, Phys. Rev. Lett. {\bf{93}}, 172502 (2004).
\bibitem{Qi0902} B. Qi, S. Q. Zhang, S. Y. Wang, J. M. Yao, and J. Meng, Phys. Rev. C, {\bf79}: 041302{R} (2009).
\bibitem{Ragnarsson88} I. Ragnarsson and P. B. Semmes, Hyperfine Interact. {\bf43}: 425 (1988).
\bibitem{Sakurai07} J. J. Sakurai, Modern Quantum Mechanics, Pearson Education (2007).



\end{thebibliography}
\end{document}